# Adiabatic physics of an exchange-coupled spin-dimer system: magnetocaloric effect, zero-point fluctuations, and possible two-dimensional universal behavior

J. Brambleby,[1, *] P. A. Goddard,[1, †] J. Singleton,[2] M. Jaime,[2] T. Lancaster,[3] L. Huang,[4]
J. Wosnitza,[4] C. V. Topping,[5] K. E. Carreiro,[6] H. E. Tran,[6] Z. E. Manson,[6] and J. L. Manson[6]

[1] *University of Warwick, Department of Physics, Gibbet Hill Road, Coventry, CV4 7AL, UK*
[2] *Los Alamos National Laboratory, National High Magnetic Field Laboratory, MS-E536, Los Alamos, NM 87545, USA*
[3] *Durham University, Centre for Materials Physics, South Road, Durham DH1 3LE, UK*
[4] *Hochfeld-Magnetlabor Dresden (HLD-EMFL), Helmholtz-Zentrum Dresden Rossendorf, D-01314 Dresden, Germany*
[5] *University of Oxford, Department of Physics, Clarendon Laboratory, Parks Road, Oxford, OX1 3PU, UK*
[6] *Eastern Washington University, Department of Chemistry and Biochemistry, Cheney, WA 99004, USA*

We present the magnetic and thermal properties of the bosonic-superfluid phase in a spin-dimer network using both quasistatic and rapidly-changing pulsed magnetic fields. The entropy derived from a heat-capacity study reveals that the pulsed-field measurements are strongly adiabatic in nature and are responsible for the onset of a significant magnetocaloric effect (MCE). In contrast to previous predictions we show that the MCE is not just confined to the critical regions, but occurs for all fields greater than zero at sufficiently low temperatures. We explain the MCE using a model of the thermal occupation of exchange-coupled dimer spin-states and highlight that failure to take this effect into account inevitably leads to incorrect interpretations of experimental results. In addition, the heat capacity in our material is suggestive of an extraordinary contribution from zero-point fluctuations and appears to indicate universal behavior with different critical exponents at the two field-induced critical points. The data at the upper critical point, combined with the layered structure of the system, are consistent with a two-dimensional nature of spin excitations in the system.

## I. INTRODUCTION

Cooperative phenomena[1–4] found in materials exhibiting a quantum critical point (QCP) (that is, a zero-temperature phase transition[5]) continue to excite great interest. Notably, spin-gapped antiferromagnetic (AF) insulators (such as $XY$-like spin $s = 1$ chains[6–8] and exchange-coupled $s = 1/2$ dimers[4,9,10]) provide examples of magnetic systems whose low-energy physics may be described in terms of a Bose-Einstein condensate (BEC) that forms between two QCPs that are reached by sweeping magnetic field[4]. The first QCP involves the closing of a spin gap where spin-1 bosons begin to condense from a non-magnetic vacuum, while the second involves a transition to a state where the number of bosons saturates. In many magnetic BEC materials, the large energy scale of the interactions means that the QCPs are most readily reached using pulsed magnetic fields, while in certain cases the field-range accessible by current technologies can limit studies to the first QCP alone (e.g. TlCuCl$_3$ [4]). These considerations necessitate a detailed experimental understanding of how these systems respond to rapid changes in magnetic field, which has so far been lacking.

While previous studies suggest that a strong magnetocaloric effect (MCE)[11,12] (a change in sample temperature $T$ on sweeping magnetic field $B$) is expected to occur at the closing of the spin-gap in pulsed fields[4], we present experimental evidence supported by a general thermodynamic model that shows that a continuous and strong MCE in fact occurs for all $B > 0$. With this insight, our results imply that a naive acceptance of the apparent sample temperature in pulsed-field measurements may lead to the results on BEC systems being interpreted erroneously[13,14]. We demonstrate this by exploiting the low-field QCPs of an organic spin-dimer network, Cu(pyz)(gly)ClO$_4$ (pyz = pyrazine; gly = glycinate)[15], to measure both pulsed-field and quasistatic magnetometry across the full phase diagram of the material. From this, we reveal an apparent discrepancy between the results of these measurements. The entropy ($S_{\rm mag}$) derived from heat capacity indicates this discrepancy is resolved if the pulsed-field measurements are taken to be approximately adiabatic. This assumption leads to the prediction of a strong MCE in the pulsed-field experiments, which we confirm via direct measurements. The heat capacity further suggests that zero-point fluctuations also contribute to $S_{\rm mag}$ and the MCE at low fields in Cu(pyz)(gly)ClO$_4$. This evidence for quantum phenomena went unreported in a previous magnetic study[15]. Finally, we find that the heat capacity exhibits different critical exponents at each QCP. Our analysis of the exponent at the upper QCP, the exchange pathways implied by the crystal structure and the results of simulations of the heat capacity are all consistent with a system that closely approximates a two-dimensional (2D) spin-dimer network, in contrast to the three-dimensional magnetic description previously suggested for this material[15]. However, from the experimental thermodynamic evidence presented here we cannot rule out the presence of an $XY$-symmetry breaking perturbation (such as a Dzyaloshinskii-Moriya interaction). Such a term would induce a crossover into a different universality class and require a different interpre-tation of the critical exponents.

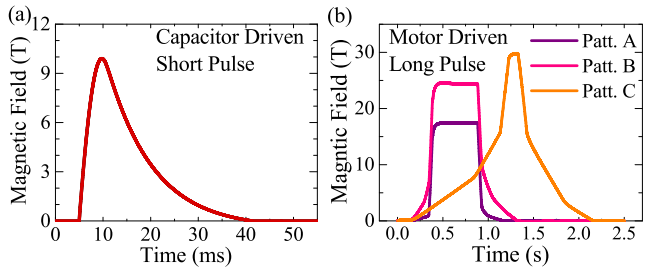

FIG. 1. Magnetic-field profiles for: (a) the capacitor driven short-pulse magnet; and (b) motor/generator driven long-pulse magnet at the National High Magnetic Field Laboratory, Los Alamos.

## II. EXPERIMENTAL DETAILS

### A. Synthesis

Single crystal samples of Cu(pyz)(gly)ClO$_4$ were prepared according to the procedure established in Ref. 15 (and the supporting information therein).

### B. Dynamic Susceptibility

*Ac Susceptibility:* Ac-susceptibility measurements were performed with a Quantum Design Magnetic Property Measurement System (MPMS) using an 8.76 mg sample. The sample is placed inside a gelatin capsule and held in position with a small amount of Vaseline to prevent the material from moving during a measurement. The magnetic moment of the sample ($M$) is measured in response to a sinusoidally oscillating field of amplitude $B_{ac}$ = 0.1 mT, held at a fixed frequency ($f_{ac}$). The component of the magnetic moment, which varies in-phase ($M'$) and out-of-phase ($M''$) with the ac-field, is recorded for quasistatic fields in the range $0 \leq B \leq 5$ T applied using a superconducting magnet. Measurements were repeated for $f_{ac}$ = 3, 33 and 99 Hz. The in-phase ($\chi'$) and out-of-phase ($\chi''$) components of the ac molar susceptibility can be derived from the measurement using $\chi^i = M^i/nB_{ac}$, where $n$ is the number of moles of the sample.

*Quasistatic proximity detector oscillator technique:* The details of the experiment have been previously reported[15] and are included here to offer new physical insight in the interpretation of analogous pulsed-field measurements below. The sample is wrapped in a Cu-coil and used as the inductive part of an LCR circuit[16]. A (MHz) radio-frequency oscillatory current is applied to the sample coil using an inductively coupled to a proximity detector oscillator (PDO). The high-frequency output from the PDO is amplified and mixed down by a two-stage heterodyne process to reduce the noise on the measure-ment. With the sample submerged in liquid helium, a quasistatic field is applied with a superconducting magnet and the resonant frequency of the LCR circuit ($\omega$) is recorded as a function of applied field. For this magnetic insulating sample, the change in $\omega$ is given by[17]:

$$\Delta\omega = -a\Delta\chi - b\Delta R_0, \quad (1)$$

where $a, b \geq 0$ are constants, $\chi$ is the susceptibility ($\mathrm{d}M/\mathrm{d}B$) and $R_0$ is the combined magnetoresistance of the sensor coil and a coax cable feeding the ac-field to the sample. The susceptibility is determined by eliminating the magnetoresistance term with a background measurement of an empty coil.

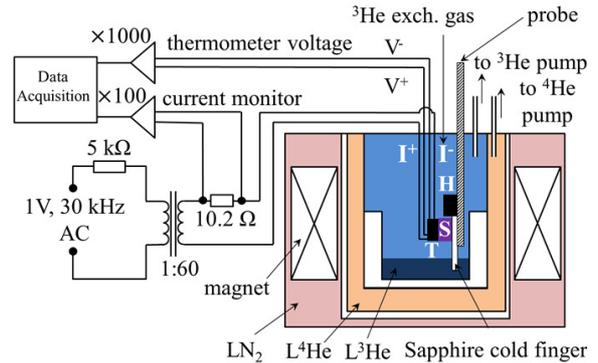

FIG. 2. (a) Schematic experimental set-up for the measurement of the MCE effect. A Cernox™ thermometer (T) is attached to the sample (S) with vacuum grease. The probe is lowered into a dual-walled $^3$He cryostat and centred in the magnet coils. The $^3$He level is tuned with a heater (H) so that a sapphire cold-finger, on to which the sample is mounted, extends into the liquid.

*Pulsed-field proximity detector oscillator technique:* A single crystal sample is loaded into a 1.3 mm diameter polychlorotrifluoroethene (PCTFE) ampoule and secured with a small amount of vacuum grease. The ampoule is wrapped in a 5-turn Cu-coil and used as the inductive part of an LCR circuit as described for the quasistatic measurement. The sample ampoule is immersed in cryogens and a capacitor driven short-pulse magnet at the National High Magnetic Field Laboratory (NHMFL), Los Alamos, provides the applied field [$B$–$t$ profile shown in Fig. 1(a)]. The magnetoresistive contribution to the change in resonant frequency can be modelled using a quadratic function. This is subtracted from the measurement of the circuit's resonant frequency (Eq. 1) to extract the sample $\chi$.

*Pulsed-field extraction magnetometry:* Pulsed-field measurements of $M$ for $B \leq 10$ T were measured with a capacitor-bank powered, short-pulse magnet. Here, the sample is placed into a PCTFE ampoule of diameter 1.3 mm and sealed with vacuum grease to prevent the sample from moving. The filled ampoule is lowered into a 1.5 mm bore, 1500-turn compensated-coil susceptometer, which is 1.5 mm in length and made from high-purity

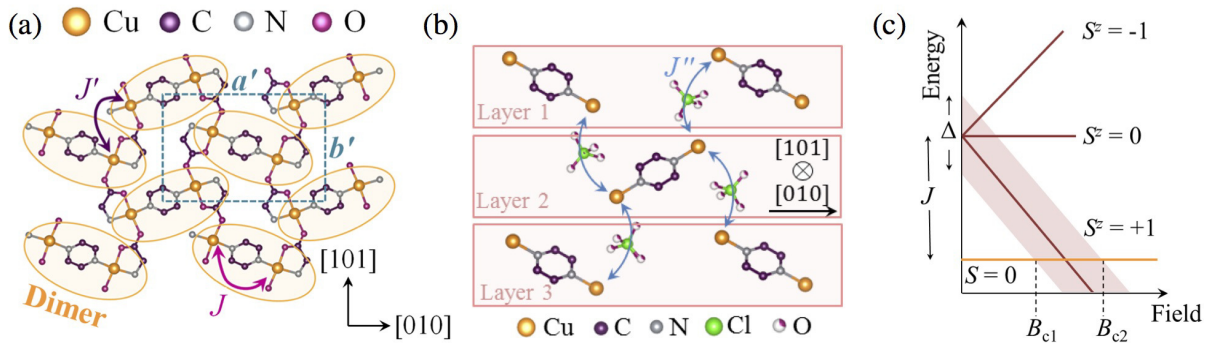

FIG. 3. (a) Cu(pyz)(gly)ClO$_4$ (pyz = pyrazine, gly = glycinate, structure measured at $T = 298$ K taken from Ref. 15) consists of corrugated sheets of dimers. Arrows mark the intradimer ($J$) and interdimer ($J'$) spin-exchange pathways. H atoms and the ClO$_4^-$ ions on sites between sheets are omitted for clarity. (b) Cu$^{2+}$ ions in adjacent planes are linked with ClO$_4^-$ anions (mediating an interlayer exchange, $J''$). The gly groups and H atoms are omitted. (c) Schematic energy-level diagram for each dimer.

Cu wire. With the sample at the center of the coil, the voltage induced in the coil as the field is pulsed is proportional to the rate of change of $M$. Numerically integrating this measurement yields the magnetization of the loaded coil. By subtracting the magnetization of an empty coil, measured under the same thermal conditions, the $M$ of the sample is deduced[18].

In all the pulsed-field measurements (magnetometry and magnetocaloric effect) the value of the magnetic field is found from integrating the voltage induced in a 10-turn coil placed at field center. The field is calibrated in each case against the copper de Haas-van Alphen oscillations (the belly orbit) induced in the signal coil of the extraction magnetometer.

### C. Magnetocaloric Effect

Measurements of the magnetocaloric effect (MCE) are performed with the 60 T long-pulse magnet (and the 65 T short-pulse magnet) at NHMFL [Fig. 2(a)]. A digital lock-in provides a constant amplitude current by feeding the 30 kHz (60 kHz) frequency, 1 V (2 V) amplitude ac voltage output through a 5 k$\Omega$ (1 k$\Omega$) resistor prior to a 1:60 turn step-up transformer. A 10.2 $\Omega$ shunt resistor following the transformer monitors the current to ensure a constant amplitude is maintained. The ac current is then passed through a Cernox™ resistor (attached to the sample) via twisted pair contacts that run approximately parallel to the applied field. The voltage drop across the Cernox™ is extracted with a second pair of contacts also wound as a twisted pair. Both the current monitor and thermometer voltages are amplified (by 100 and 1000 times respectively) using a Stanford Research Systems low noise preamplifier and recorded as a function of time during the pulse with the lock-in amplifier set to record at the same frequency as the output.

The Cernox™ is mounted directly on the sample with vacuum grease, which is in turn attached to insulating cigarette paper with a further application of grease. To promote adiabatic conditions, the paper is glued to a sapphire cold finger with GE varnish so that the sample is positioned $\approx 10$ mm above the bottom of a He-cryostat. In this configuration, the cold finger extends into liquid He that collects in the cryostat tail, while the sample itself is in gas. The liquid-He level is tuned prior to the field sweeps by pulsing a heater close to the sample, ensuring the sample was out of the liquid.

During each field pulse ($B$–$t$ profiles shown in Fig. 1), the in-phase and out-of-phase components of the thermometer ac-current and voltage are recorded and the phase of the signal is defined to minimise the out-of-phase component of the current monitor. The ratio of the in-phase voltage to the in-phase current amplitude is used as the measure of the resistance of the Cernox™.

The magnetoresistance ($R$) of the Cernox™ is calibrated with no sample in place using the same pulse-patterns starting from initial temperatures in the range $0.45 \leq T \leq 10$ K. The $B$-dependence of $R$ is modelled with the invertible empirical relation:

$$R = \exp\left(a_1 + \frac{a_2}{T^{a_3}}\right), \quad (2)$$

where the coefficients $a_i$ are represented as polynomials in $B$:

$$a_i = \sum_{j=0}^{4} c_{ij} B^j. \quad (3)$$

Here, $a_3$ is constrained to be linear in $B$ in order to uniquely determine a complete set of coefficients for $a_{1,2}$. Within this model, the thermometer $T$ can be deduced from a measurement of $R(B)$. The error in $T$ extracted from this method is 1–3% for the $T, B$ range of interest.

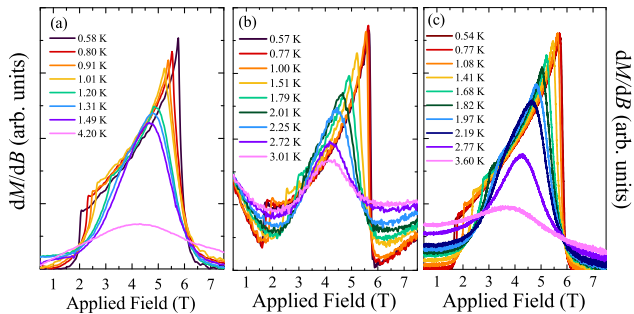

FIG. 4. (a) $dM/dB$ vs. quasistatic magnetic field for Cu(pyz)(gly)ClO$_4$ from PDO experiments[15]. The asymmetry of $dM/dB$ points to the influence of zero-point fluctuations at $B_{c1}$. This is discussed following the heat capacity analysis below. (b) PDO susceptometer and (c) extraction magnetometer measurements of $dM/dB$ (this work) in pulsed fields. Data are labeled by helium-bath temperature $T_0$.

## III. RESULTS AND DISCUSSION

### A. Arrangement of dimers

In Cu(pyz)(gly)ClO$_4$, the Heisenberg $s = 1/2$ Cu$^{2+}$ ions form dimers where pyz molecules mediate an AF intradimer exchange ($J$) [15]. An interdimer exchange ($J'$) is provided by gly groups that ligate each dimer to four neighbors to form a corrugated sheet as shown in Fig. 3(a). The dimer centres are coplanar and individual layers stack along the [10$\bar{1}$] direction. Each plane is displaced relative to the sheet below along the [101] vector. Perchlorate anions (ClO$_4^-$) connect the Cu$^{2+}$ ions between each adjacent layer. In Fig. 3(b) the one-third filled spheres for the O atoms denote a partial site-occupation of three equally likely positions and the Cl-O bond lengths range from 1.395−1.448 Å at 298 K. The stacking of dimer layers results in Cu···O-Cl-O···Cu pathways that provide the interlayer exchange, $J''$. The total distance along these non-linear pathways is ≈ 8.3 Å, while the linear interplanar nearest-neighbour Cu-Cu distance is 6.6206(9) Å. A full description of the structural refinement from x-ray diffraction is available from Ref. 15.

The longer axis of the distorted CuN$_2$O$_4$ octahedra is the Cu-O coordination bond to two separate ClO$_4^-$ ions above and below the sheets. This axial elongation of the coordination environment constrains the $d_{x^2-y^2}$ orbital of the unpaired Cu$^{2+}$ electron to lie predominantly within the pyz-gly planes. The resulting low spin density expected between the layers therefore suggests that the dominant exchange pathways will be between coplanar Cu$^{2+}$ ions.

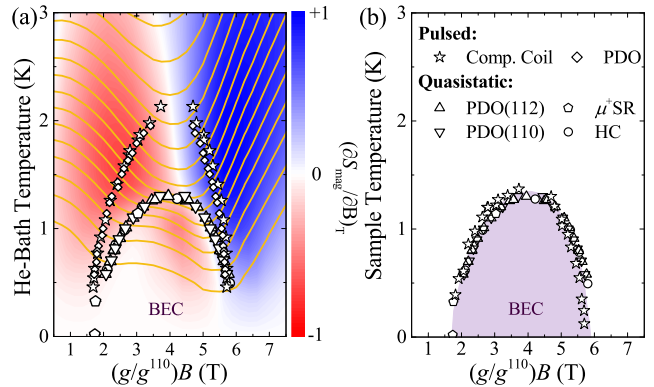

FIG. 5. (a) $B-T$ phase diagram of Cu(pyz)(gly)ClO$_4$. Quasistatic measurements [heat capacity (HC); PDO and $\mu^+$SR] map out a dome enclosing a BEC bounded by a low-field disordered and high-field saturated phase[15]. Diamonds and stars mark peaks in $dM/dB$ from pulsed fields [Fig. 4(b) and (c)] plotted vs. $T_0$. Lines are experimental isentropes (0.2 JK$^{-1}$mol$^{-1}$ intervals) derived from heat capacity. Colormap = $(\partial S_{\text{mag}}/\partial B)_T$ (arb. units). (b) The same data when the pulsed-field measurements are assumed adiabatic and the sample temperature is inferred from the isentropes. As in Ref. [15], the field axis in both cases is scaled by the experimental $g$-factor divided by that measured normal to the (110) plane, so that results from different measurements in which the field is applied along different directions can be compared.

### B. Mapping phase boundaries using magnetometry and heat capacity

Antiferromagnetically-coupled $s = 1/2$ dimers (with total spin, $S$) have a singlet ($S = 0$) ground-state, a triplet of excited states, and a gap between the two that can be closed by a magnetic field [see the schematic in Fig. 3(c)]. For dimer networks with uniaxial symmetry, a Matsubara-Matsuda transformation of the spin Hamiltonian maps the lowest lying states in non-zero applied field onto a hard-core boson picture where each dimer is represented by the $S = 0$ or $S^z = +1$ (triplon) state[4,19,20]. A finite $J'$ permits triplons to hop to neighboring sites[19] giving the excitations a dispersion

$$E(\mathbf{k}) = J + 2J' \cos k_x \cos k_y - g\mu_B B, \quad (4)$$

where $\mathbf{k} = (k_x, k_y)$ is the dimensionless triplon wavevector. The excited states have a bandwidth $\Delta = 4J'$, hence the spin gap closes over a range of $B$ bounded by two QCPs that obey the relation $g\mu_B B_{c1,2} = J \mp 2|J'|$.

For the temperature range across which Cu(pyz)(gly)ClO$_4$ has been studied, no low-energy symmetry-breaking perturbations have so far been detected. It is known that such perturbations (e.g. a Dzyaloshinskii-Moriya interaction) begin to modify the universal behaviour only at temperatures less than the energy scale of the perturbation[4]. Above this temperature and below the thermodynamic limit the system can be described within the boson description. For the analysis that follows we assume that our material maps onto the boson model described by Eq. 4. While we cannot absolutely rule out the presence of an $XY$-symmetry breaking perturbation, as will be seen, the simulations of the thermodynamic properties arising from the boson model agree extremely well with the experimental data.

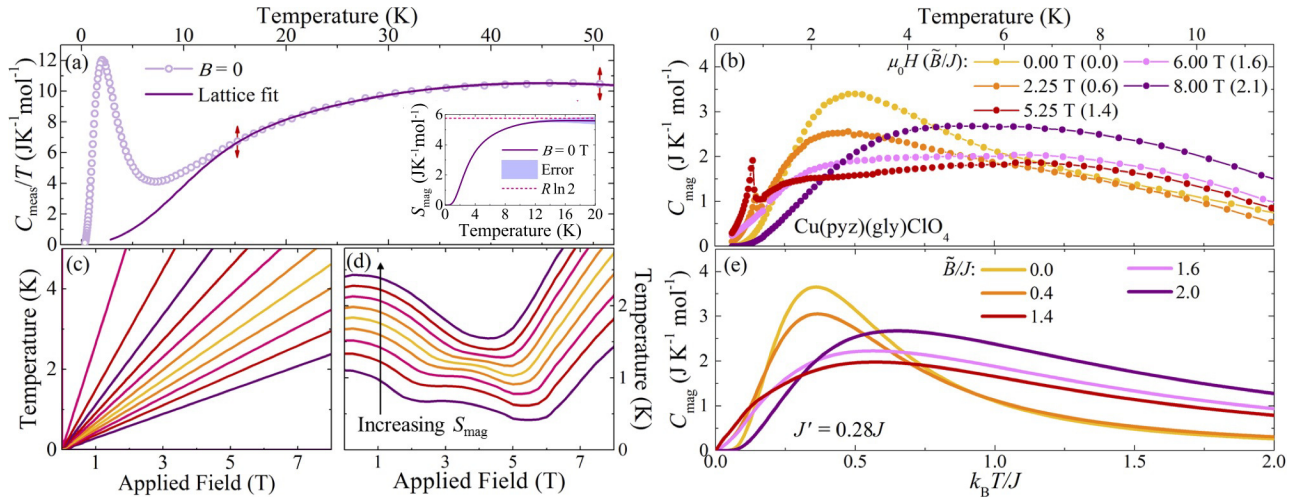

FIG. 6. (a) $C_{\text{meas}}/T$ vs. $T$ for Cu(pyz)(gly)ClO$_4$ at zero field (circles)[15]. Data for $15 \leq T \leq 50$ K are modelled by one Debye and one Einstein mode (line)[21]. The inset shows that the zero-field entropy change per Cu$^{2+}$ is consistent with $R\ln 2$, i.e. $R\ln 4$ per dimer, as required for AF $s = 1/2$ dimers. (b) $C_{\text{mag}}$ vs. $T$ (this work) for $0 \leq \mu_0 H \leq 8$ T (labels of $\tilde{B}/J = g\mu_B B/J$ are shown in brackets to aid comparison with the simulations). (c) Calculated isentropes for an ideal $s = 1/2$ paramagnet. (d) Experimental isentropes for Cu(pyz)(gly)ClO$_4$. (See Appendix A.) (e) Simulated heat capacity for a system of exchange-coupled dimers with a two-dimensional dispersion and $J' = 0.28J$.

The positions of the QCPs in Cu(pyz)(gly)ClO$_4$ are determined from measurements performed isothermally in the quasistatic fields supplied by a superconducting magnet. Published[15] susceptibility ($\chi = \text{d}M/\text{d}B$; $M$ = magnetization) measurements employing the proximity-detector oscillator (PDO) technique show sharp kinks at the field-induced phase transitions for helium-bath temperatures ($T_0$) below 1.2 K [see Fig. 4(a)]. These data, plus quasistatic heat capacity and muon-spin rotation ($\mu^+$SR) measurements[15], map out a dome in the $B-T$ plane enclosing the BEC phase, as shown in Fig. 5(a) by triangles, pentagons and circles. The QCP positions, $g\mu_B B_{c1}/k_B = 2.5(1)$ K and $g\mu_B B_{c2}/k_B = 9.0(2)$ K [15], together imply that $J/k_B = 5.8(3)$ K and $|J'|/k_B = 1.6(1)$ K. However, measurements of $\text{d}M/\text{d}B$ using both a PDO susceptometer and an extraction magnetometer repeated in rapidly-changing pulsed magnetic fields [pulse height 10 T, rise time $\approx 10$ ms, Fig. 1(a)] show kinks, analogous to the quasistatic measurements, but which persist for $T_0$ values in excess of 1.8 K [see Fig. 4(b) and (c)]. When mapped onto the phase diagram in Fig. 5(a), these points appear to form a dome that is extended by comparison with the results of quasistatic experiments.

On first viewing, the apparently extended dome measured using pulsed-fields in our material resembles that obtained from pulsed-field ultrasound and magnetometry measurements on Sr$_3$Cr$_2$O$_8$, a dimer system where critical fields vs. $T_0$ are reported to map out a magnon liquid state that extends above a low-temperature ordered phase (akin to a BEC)[13,14]. However, in our system (and by extension related materials) we will present heat-capacity measurements that strongly suggest that the apparent extension of the low-temperature dome in pulsed-field experiments should instead be attributed to a MCE brought about by approximately adiabatic rapid field sweeps, removing the need to invoke a separate magnetic phase to explain the results.

The $B = 0$ heat capacity[15] of Cu(pyz)(gly)ClO$_4$ ($C_{\text{meas}}$) shown in Fig. 6(a) exhibits a broad maximum due to a contribution from the Cu$^{2+}$ ions ($C_{\text{mag}}$) superimposed on a sloping background from lattice vibrations ($C_{\text{latt}}$). Assuming $C_{\text{latt}}(T)$ to be independent of $B$, we subtract it from the data collected at each value of applied field[21]. The resultant $C_{\text{mag}}(T)$ data are shown in Fig. 6(b) and exhibit two distinct features : a $\lambda$-peak at $T_{\text{BEC}}$ (for $B_{c1} \leq B \leq B_{c2}$) marking the BEC phase; and $B$-dependent broad maxima at temperatures above $T_{\text{BEC}}$.

The magnetic entropy ($S_{\text{mag}}$) of Cu(pyz)(gly)ClO$_4$ is found by integrating $(C_{\text{mag}}/T)|_B$ (constant $B$) for $T \leq 10$ K, taking $C_{\text{mag}} = 0$ at $T = 0$. By compiling $S_{\text{mag}}(T)|_B$ curves, the experimental entropy in the $B-T$ plane can be determined (see Appendix A). Isentropes (lines of constant entropy) indicate the path that will be taken across a phase diagram on sweeping magnetic field adiabatically. For example, in the case of a paramagnetic (PM) ensemble of spins the partition function and $S_{\text{mag}}$ are functions of $B/T$, which remains constant in an adiabatic field sweep[23]; hence the isentropes are straight lines as shown in Fig. 6(c). The isentropes of Cu(gly)(pyz)ClO$_4$ shown Fig. 6(d) thus describe how the sample temperature evolves in an adiabatic field sweep. Two local minima in the isentropes at low temperatures mark the field positions of the QCPs. Here, the competition of magnetic phases leads to a locally high entropy so the isentropes drive the sample temperature lower to maintain constant $S_{\mathrm{mag}}$. For $B > B_{\mathrm{C2}}$ the isentropes become linearly dependent on field since the Zeeman energy dominates spin-exchange couplings and $S_{\mathrm{mag}} \approx S_{\mathrm{mag}}(B/T)$.

The discrepancy between the results of pulsed and quasistatic magnetometry is then resolved if the sample temperature is assumed constant in quasistatic measurements, while pulsed-field experiments are assumed to be close to adiabatic, such that the sample temperature is inferred by following an isentrope [reproduced in Fig. 5(a)]. With this correction, the features in the pulsed-field data map directly onto the small dome that encloses the BEC phase as measured under quasistatic conditions [see Fig. 5(b)]. This agreement in the positions of features observed across the different measurement techniques therefore: (i) strongly supports the adiabatic nature of the pulsed-field data, (ii) implies that a significant MCE ($|\Delta T|/T_0 \leq 0.37$) occurs during the field pulses, and (iii) shows that the pulsed-field measurements are sensitive to the same low-temperature phase measured in quasistatic experiments.

We note that data from ac-susceptibility measurements (not shown) in quasistatic $B \leq 5$ T with various ac frequencies $\leq 99$ Hz all agree with those from PDO measurements in quasistatic fields. This indicates that isothermal and adiabatic conditions are dictated by the applied field sweep rate rather than the ac-field modulation frequency ($\approx 20$ MHz in PDO experiments).

### C. Measuring the magnetocaloric effect

The MCE was investigated directly by attaching a calibrated Cernox™ resistor to a single-crystal sample and monitoring its temperature ($T_{\mathrm{Cer}}$) during pulsed-field sweeps using a 4-wire resistance technique. The magnetic field-sweep rate was varied using the $B$–$t$ profiles (shown in Fig. 1) of the capacitor-driven short-pulse (SP) and motor/generator-driven long-pulse (LP) magnets at the National High Magnetic Field Laboratory, Los Alamos. The SP and LP MCE measurements will likely have a different degree of adiabaticity to each other (and to the SP magnetometry experiments). Nevertheless, the isentropes derived from the heat capacity data suggest that any experiment which departs from ideal isothermal conditions should exhibit an MCE as the field is swept.

The temperatures traced out with SP down sweeps shown in Fig. 7(a) resemble the isentropes of Fig. 6(d). A kink in $T_{\mathrm{Cer}}$ close to 2 T denotes a phase transition and a linear thermal response above 5.5 T signifies a switch to the saturated (spin-polarised) state. The low-field kink in $T_{\mathrm{Cer}}$ coincides with a change in sign of $\Gamma = T^{-1}(\partial T/\partial B)$ (for $S_{\mathrm{mag}} \approx$ constant) and is located by finding a peak in d$\Gamma$/d$B$ [24]. This enables the lower critical field to be determined independent of linear background contributions to $T_{\mathrm{Cer}}$. The upper boundary is defined by the point where $T_{\mathrm{Cer}}(B)$ departs from a linear fit to the data for $B \geq 7$ T (dashed lines). See Appendix B for more details on how these features are determined.

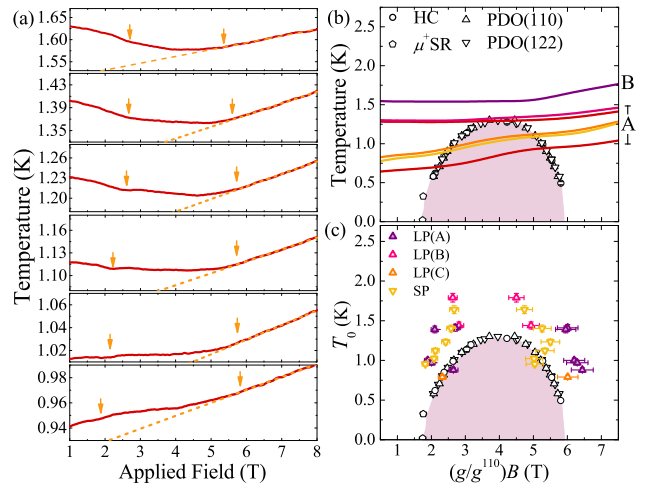

FIG. 7. (a) $T_{\mathrm{Cer}}$ vs. applied field from SP downsweeps. Arrows mark critical fields (see text). (b) LP measurements of the MCE (lines) superimposed on the $B$-$T$ phase diagram. Labels denote the field pattern used as in Fig. 1(b). (c) The same $B$-$T$ phase diagram with critical fields from MCE measurements (triangles) plotted against $T_0$.

The measured temperature shift ($\Delta T_{\mathrm{Cer}}$) is found to be maximised when d$B$/d$t$ is limited to $\approx 100$ Ts$^{-1}$ with the LP magnet. The MCE data in Fig. 7(b) clearly show the development of a double minimum in $T_{\mathrm{Cer}}$ on reducing $T_0$ and sweeping past the two QCPs. Critical fields were determined from peaks in d$\Gamma$/d$B$ and are plotted against $T_0$ as triangles in Fig. 7(c). The peaks follow an apparent extended dome, similar to that mapped out from pulsed-field magnetometry, thus supporting the conclusion that a MCE occurs in these measurements and gives rise to the observed phase transitions, even when $T_0$ exceeds the top of the BEC phase. It is this observation that we discuss in the next section.

### D. Explaining the magnetocaloric effect by simulating the entropy of exchange-coupled dimers

We explain the MCE in the dimer system with a model considering the thermal occupation of bands of triplon states in an applied magnetic field. The heat capacity of independent dimers is deduced from the partition function and the interdimer interactions are accounted for by considering a two-dimensional dispersion of the triplon states. The full details of how the simulations are performed can be found in Appendix C, while the dimen-sionality of the real system is discussed in the following section.

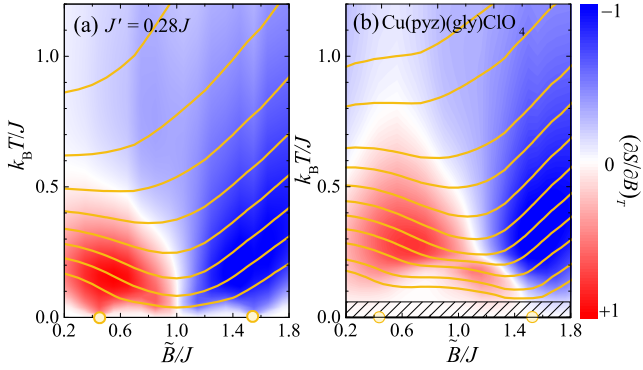

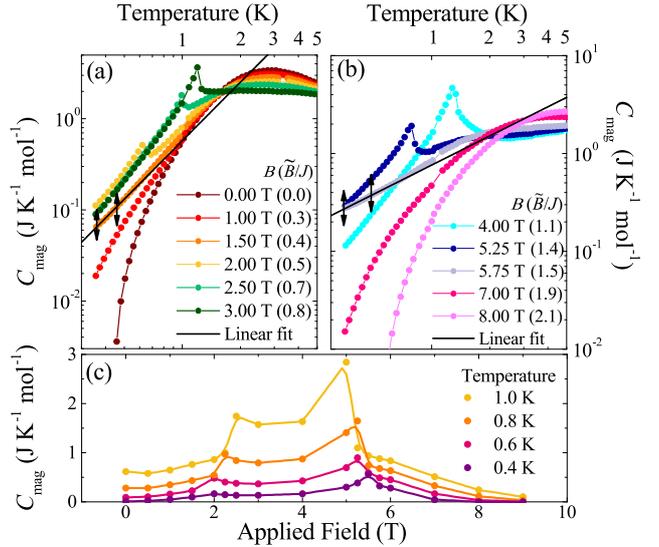

FIG. 8. (a) Simulated isentropes in the $\tilde{B}$-$T$ plane ($\tilde{B} = g\mu_B B$) for coupled dimers with a two-dimensional dispersion ($J' = 0.28J$). (b) Experimentally determined isentropes for Cu(pyz)(gly)ClO$_4$. Colormap = $(\partial S/\partial \tilde{B})_T$ (arb. units); yellow dots show $B_{c1,2}$; hatched region marks a region below the minimum accessible $T$ in the $C(T)$ measurements.

FIG. 9. $C_{\mathrm{mag}}$ vs. $T$ for Cu(pyz)(gly)ClO$_4$ (points) plotted on a logarithmic scale. Data are labelled by $B$ (and $\tilde{B}/J = g\mu_B B/J$) for (a) $\tilde{B}/J \leq 0.8$; (b) $\tilde{B}/J \geq 1.1$. Data at 1.50 and 5.75 T are modelled by $C_{\mathrm{mag}} \propto T^n$ for $T \leq 0.52$ K (solid lines). (c) $C_{\mathrm{mag}}(B)$ for Cu(pyz)(gly)ClO$_4$ (points) from constant $T$ cuts through $C(T)|_B$ data. Lines = linear interpolation of the points, smoothed over a 0.3 T window.

With $J' = 0.28J$ (as expected for Cu(pyz)(gly)ClO$_4$ from the positions of the critical fields), broad maxima in the simulated heat capacity [Fig. 6(e)] share features with the measured data [Fig. 6(b)]. While the simulation cannot replicate the $\lambda$-peaks bounding the BEC phase in the real system, the broad features of both datasets may be explained as Schottky anomalies resulting from the field-splitting of dimer spin-states. For $\tilde{B} \ll J$ ($\tilde{B} = g\mu_B B$), a peak in $C_{\mathrm{mag}}(T)$ [Fig. 6(e), yellow line] results from the depopulation of closely spaced triplon levels to a singlet state. This maximum is suppressed in height when $\tilde{B}$ is raised (orange) as the gap between the $S^z = +1$ and $S = 0$ levels decreases. For $\tilde{B} \approx J$ (red) a low-temperature shoulder develops below the main Schottky anomaly due to thermal depopulation within the two lowest-lying spin states. The high-temperature maximum seen at this value of $\tilde{B}$ is sensitive to the occupation of excited states relative to these low-lying levels. For $\tilde{B} \geq 1.5J$, the temperature of the broad peak tends towards a linear dependence on $\tilde{B}$ (pink, purple). This behavior (analogous to a PM spin system) occurs in dimers for $\tilde{B} > \tilde{B}_{c2}$ where the zero-field splitting of levels becomes decreasingly relevant to the occupation of the spin states.

Reconstructing the entropy from the simulated heat capacity [Fig. 8(a)], we predict that areas of large $|(\partial S/\partial \tilde{B})_T|$, and hence a strong MCE, occur in adiabatic field-sweeps at temperatures above each QCP (yellow dots), as observed in the real material [Fig. 8(b)]. Comparing the isentropes (lines) against the measured data, we note that for $k_B T_0 \approx J$ no cooling is expected on sweeping $B$ adiabatically. As $T_0$ is lowered, cooling is immediately induced for $B > 0$ and not just at $B_{c1}$ as has been previously suggested[4]. We also confirm that an analogous effect occurs on approaching $B_{c2}$ from the high-field side, a field regime that remains untested in many inorganic BEC systems where $B_{c2}$ can be inaccessibly large[4]. For $\tilde{B} > \tilde{B}_{c2}$, the isentropes are linear and have a smaller gradient at lower temperatures, as in PM systems. It is thus clear that a MCE in adiabatic or quasi-adiabatic measurements of spin-gapped materials must be carefully considered at all fields when interpreting the sample temperature.

The effects of correlations in the real system (e.g. field-induced symmetry breaking at $B_{c1}$) leads to differences between the experimental data and the calculations [Fig. 8(a) and (b)]. These include the minima in the measured isentropes at $B_{c1,2}$ when $k_B T \ll J$. Furthermore, we find the sample cooling from the MCE persists to higher temperatures in the measured data than in the simulations. This difference is not reconciled when an interlayer triplon dispersion is included in the calculation (see Appendix C), suggesting there are additional contributions to $S_{\mathrm{mag}}$ not accounted for by our simple thermodynamic model.

### E. Determining dimensionality and the effect of zero-point fluctuations

Evidence that zero-point fluctuations influence $C_{\mathrm{mag}}(T)$ and the MCE in Cu(pyz)(gly)ClO$_4$ is found by plotting the heat capacity as a function of $T$ on a logarithmic scale as shown in Fig. 9(a) and (b). Here it can be seen that $C_{\mathrm{mag}} \propto T^n$ only when the applied field is close to either of the two critical fields, which occur when $B \approx 1.5$ T and 5.8 T in the heat capacity measurements[15] ($\tilde{B}/J = 0.4$ and 1.5). A universal power-law is expected of BEC systems when $B = B_{c1,2}$ where the critical exponent $n = d/2$ reflects the dimensionality ($d$) of the system[4,25]. Considering only the $C_{\text{mag}}(T)$ data below 40% of the maximum ordering temperature (where it is known that the most reliable exponents are obtained[26]), the traces at $\tilde{B}/J = 0.4$ and 1.5 yield $n = 2.13(2)$ and $0.99(2)$ at $B_{c1}$ and $B_{c2}$, respectively. The value of the critical exponent at $B_{c2}$ (the exponent at $B_{c1}$ is discussed below) implies that Cu(pyz)(gly)ClO$_4$ has strong correlations only within the dimer layers, which is in contrast to the suggestion of three-dimensional magnetic behaviour previously given for this system[15], but confirms the expectations resulting from the discussion of the structure in Section III A.

Ideal two-dimensional AF $s = 1/2$ dimer systems are predicted[9,10] to exhibit a low-temperature Berezinskii-Kosterlitz-Thouless (BKT) phase[27,28]. However, crossing a BKT phase boundary is not expected to yield any signatures in probes such as heat capacity[9] or magnetometry[10]. Our measurements do map out a low-temperature ordered phase in Cu(pyz)(gly)ClO$_4$, which therefore suggests that the interlayer interactions in our material, while small, are finite. Further evidence of the anisotropy comes from a comparison of the experimental and simulated heat capacity using a three-dimensional dispersion (Appendix C), which also indicates that the interlayer interaction is very much smaller than the interdimer interaction within the layers, corroborating the quasi-two-dimensional nature of the triplon excitations in this material.

Isothermal cuts through the $C_{\text{mag}}(T)|_B$ data [Fig. 9(c)] show that maxima in $C_{\text{mag}}(B)$ bounding the BEC phase are smaller at $B_{c1}$ than at $B_{c2}$. This asymmetry (also evident in d$M$/d$B$, Fig. 4) resembles the published[29] heat capacity of the $s = 1$ BEC system NiCl$_2$-4SC(NH$_2$)$_2$ (DTN). In that material, the reduced size of the peak at $B_{c1}$ is attributed to its proximity to the disordered zero-field state and the large ratio for $B_{c2}/B_{c1} \approx 6$. In the low-field disordered phase of this system zero-point fluctuations renormalise the effective mass of bosonic excitations, while this is not expected to occur close to the saturated (spin-polarized) state where $M$ is a conserved quantity[29]. The asymmetry we observe in Fig. 9(c) therefore suggests the renormalization effects of zero-point fluctuations are also strong at $B_{c1}$ in $C_{\text{mag}}(B)$ in Cu(pyz)(gly)ClO$_4$.

It is possible that the zero-point fluctuations at $B_{c1}$ could also affect the temperature range over which universal behaviour can be observed at each QCP. This might account for the different exponents we find at $B_{c1}$ and $B_{c2}$ when considering $C_{\text{mag}}(T)$ over the same temperature window[30,31] and justifies the use of the critical exponent at $B_{c2}$ in determining the dimensionality of the system.

As mentioned previously, it is possible that an undetected symmetry-breaking perturbation is significant in our experimental temperature regime. In particular, it is predicted that a Dzyaloshinskii-Moriya interaction could emerge in our material at sufficiently low temperatures[15]. Such a term would induce a crossover from the BEC ($XY$) to Ising universality class. If the energy scale of this potential perturbation is indeed of the order of 400 mK or above, then it would necessitate a different interpretation of the universal behavior we observe in our measurements. In contrast to the $n = d/2$ behaviour expected for the XY model, the exponent of the heat capacity close to the critical magnetic field for Ising systems is known to be $n = d$ [30]. Thus, neither of our measured exponents are characteristic of a 3D Ising system. The exponent observed at $B_{c1}$ could be consistent with our expectation of two-dimensional behaviour in this system. However, within this scheme the fact that the exponent changes to $0.99(2)$ at $B_{c2}$ is perplexing and would require further explanation.

## IV. SUMMARY

Cu(pyz)(gly)ClO$_4$ is found to be a quasi-two-dimensional $s = 1/2$ exchange-coupled dimer system whose two magnetic field induced QCPs are within the range of both pulsed and quasistatic laboratory fields. The molecular building blocks from which the material is constructed allow for systematic chemical tuning of the structure. These features, together with the reported role played by zero-point fluctuations in the magnetic properties, suggest that a continued examination of this system would shed further light on the BEC in quantum magnets.

The above analysis of the magnetometry and heat-capacity data shows that a substantial MCE can be induced in systems with level crossings or strong correlations when studied with pulsed magnetic fields. This necessitates a careful consideration of the experimental sample temperature to avoid incorrect interpretation of results. On the other hand, because the cooling in adiabatic magnetic-field sweeps allows lower sample temperatures to be achieved, it is possible that exploiting the MCE may be advantageous in future studies of emergent phenomena near QCPs.

## V. ACKNOWLEDGMENTS

JB, TL and CVT thank EPSRC for support. PAG acknowledges that this project has received funding from the European Research Council (ERC) under the European Union's Horizon 2020 research and innovation programme (grant agreement No. 681260). A portion of this work was performed at the National High Magnetic Field Laboratory, which is supported by National Science Foundation Cooperative Agreement No. DMR-1157490 and the State of Florida, as well as the *Strongly Correlated Magnets* thrust of the DoE BES "Science in 100

T" program. CVT thanks Gavin Stenning for help on the Quantum Design PPMS instrument in the Materials Characterisation Laboratory at the ISIS Neutron and Muon Source. The work at EWU was supported by the NSF through grant no. DMR-1306158.

## Appendix A: Entropy in the $B-T$ plane

### 1. Entropy of a spin-1/2 paramagnet

For a paramagnetic (PM) Heisenberg spin ($\mathbf{s}$) in an applied magnetic ($B$) field, the single-particle Hamiltonian ($\mathcal{H}$) is governed by a Zeeman term (see e.g. Ref. 34): $\mathcal{H} = g\mu_\mathrm{B}\mathbf{B}\cdot\mathbf{s}$, where $g$ is the isotropic $g$-factor, such that the partition function ($\mathcal{Z}$) for spin $s = 1/2$ particles at temperature $T$ is given by

$$\mathcal{Z} = 2\cosh\left(\frac{g\mu_\mathrm{B}B}{2k_\mathrm{B}T}\right). \tag{A1}$$

The magnetic entropy ($S_\mathrm{mag}$) is then determined from

$$\frac{S_\mathrm{mag}}{k_\mathrm{B}} = -T\frac{\partial \ln\mathcal{Z}}{\partial T} + \ln\mathcal{Z}, \tag{A2}$$

yielding

$$\frac{S_\mathrm{mag}}{k_\mathrm{B}} = -\frac{g\mu_\mathrm{B}B}{2k_\mathrm{B}T}\tanh\left(\frac{g\mu_\mathrm{B}B}{2k_\mathrm{B}T}\right) + \ln\left[2\cosh\left(\frac{g\mu_\mathrm{B}B}{2k_\mathrm{B}T}\right)\right]. \tag{A3}$$

Here, $S_\mathrm{mag}$ is a function of the ratio $B/T$ so the sample temperature is constrained to have a linear dependence on an applied field in an adiabatic field sweep as shown in Fig. 6(c). This result is appropriate for PM spin ensembles as well as strongly-correlated systems when the energy scale of the magnetic field dominates the intrinsic spin-exchange interactions within the system. For Cu(pyz)(gly)ClO$_4$ this corresponds to fields in excess of the upper critical field, $B_\mathrm{c2}$ [Fig. 6(d)].

### 2. Entropy of Cu(pyz)(gly)ClO$_4$

The published[15] heat capacity of Cu(pyz)(gly)ClO$_4$ was recorded for seventeen constant fields in the range $0 \le B \le 9$ T. For each measurement of the heat capacity at constant $B$ [$C_\mathrm{meas}(T)|_B$], the magnetic contribution to the data was determined as described in Ref. 21 and $S_\mathrm{mag}(T)|_B$ was deduced from

$$S_\mathrm{mag}(T)|_B = \int_0^T \frac{C_\mathrm{mag}(T')|_B}{T'}dT', \tag{A4}$$

for $T \le 10$ K assuming $C_\mathrm{mag} = 0$ at $T = 0$. The field dependence of $S_\mathrm{mag}$ was inferred from constant $T$ cuts through $S_\mathrm{mag}$ in the $B-T$ plane. These data were linearly interpolated to an interval of $\Delta B = 0.25$ T and smoothed with a low-pass digital filter through the application of a 5 point moving-window average function in MATLAB[35]. The interpolated data were subsequently differentiated and smoothed (as above) to determine the form of $(\partial S_\mathrm{mag}/\partial B)_T$ vs. $B$. These data were used to construct the colormap plot in Fig. 5(a).

## Appendix B: Extracting critical fields from magnetocaloric effect data

*Short Pulse Magnet:* Minima in the Cernox™ temperature ($T_\mathrm{Cer}$), found on sweeping the field through critical regions of the $B-T$ phase diagram of Cu(pyz)(gly)ClO$_4$, coincide with a change in sign of $\Gamma = T^{-1}(\partial T/\partial B)$ [24]. A rapid change in $\Gamma$, induced by the condensation of triplons generates a peak in $\mathrm{d}\Gamma/\mathrm{d}B$ [Fig. 10(a)]. This is used as a measure of the lower-critical field. The transition to a spin-polarised phase is more subtle in $\mathrm{d}\Gamma/\mathrm{d}B$. To determine the upper critical field, a linear fit is made to the $T_\mathrm{Cer}(B)$ data above 7 T. The departure between the line and the measured data is approximately $\pm 0.3\%$ within the fitted range. The critical field $B_\mathrm{c2}$ is determined from the point when the linear fit departs by more than 0.3% from the measured $T_\mathrm{Cer}$.

*Long-pulse magnet:* Critical fields were extracted from peaks in $\mathrm{d}\Gamma/\mathrm{d}B$ [Fig. 10(b), (c), and (d) for pulse patterns A, B, and C of Fig. 1(b), respectively]. We found that limiting the magnetic field sweep rate with the long-pulse magnet increased the sensitivity to changes in the sample temperature relative to the short-pulse experiments. Hence, both the upper and lower critical fields were able to be determined from peaks in $\mathrm{d}\Gamma/\mathrm{d}B$. The location of these features in the MCE measurements were checked for consistency against $\mathrm{d}\Gamma/\mathrm{d}B$ curves found from the isentropes derived from the experimental heat capacity and a similar starting temperature [Fig. 10(e), (f), and (g)].

## Appendix C: Simulating the entropy of a spin-dimer system

### 1. Independent AF $s = 1/2$ dimers

For AF $s = 1/2$ dimers, the energy of the singlet ($E_\mathrm{s}$) and triplet ($E_\mathrm{t}$) dimer states are $E_\mathrm{s} = 0; E_\mathrm{t} = J, J \pm \tilde{B}$, where $\tilde{B} = g\mu_\mathrm{B}B$, such that the singlet-triplet gap is closed when $\tilde{B} = J$. The partition function (in terms of $\beta = 1/k_\mathrm{B}T$) is

$$\mathcal{Z} = 1 + e^{-\beta J}\left(1 + 2\cosh\beta\tilde{B}\right), \tag{C1}$$

and the energy ($U = -\partial \ln\mathcal{Z}/\partial\beta$) is given by

$$U = e^{-\beta J}\left[\frac{J\left(1 + 2\cosh\beta\tilde{B}\right) - 2\tilde{B}\sinh\beta\tilde{B}}{1 + e^{-\beta J}\left(1 + 2\cosh\beta\tilde{B}\right)}\right]. \tag{C2}$$

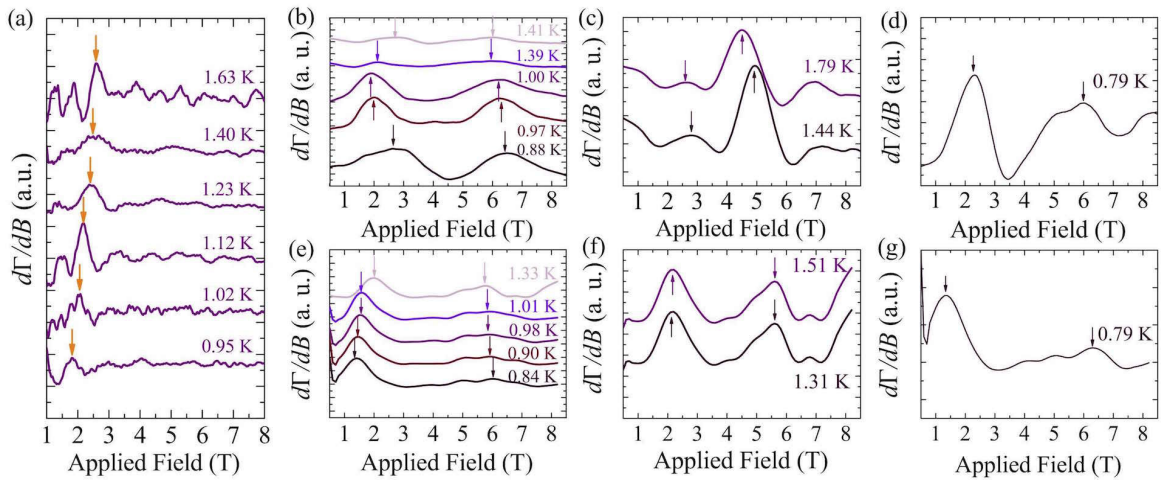

FIG. 10. (a) Peaks in $d\Gamma/dB$ (arrows) from measurements of the Cernox™ $T$ in a short-pulse magnet mark the position of the low-field phase transition in Cu(pyz)(gly)ClO$_4$. Data are offset for clarity and ordered as arranged in Fig. 7(a). Measurements in the long-pulse magnet using field patterns: (b) A; (c) B, and (d) C. These data are compared against $d\Gamma/dB$, as determined from the experimental isentropes [Fig. 6(d)], in panels (e), (f), and (g). Data are labelled by the initial $T$ prior to the field sweep and the curves are offset for clarity.

The heat capacity ($C_{\rm mag} = dU/dT$) is then

$$\frac{C_{\rm mag}}{k_{\rm B}\beta^2} = e^{-\beta J}\left[\frac{f(J,\tilde{B})}{1+e^{-\beta J}\left(1+2\cosh\beta\tilde{B}\right)}\right] \quad \text{(C3)}$$
$$-e^{-2\beta J}\left[\frac{J\left(1+2\cosh\beta\tilde{B}\right)-2\tilde{B}\sinh\beta\tilde{B}}{1+e^{-\beta J}\left(1+2\cosh\beta\tilde{B}\right)}\right]^2,$$

where

$$f(J,\tilde{B}) = J^2\left(1+2\cosh\beta\tilde{B}\right)-4\tilde{B}J\sinh\beta\tilde{B}$$
$$+2\tilde{B}^2\cosh\beta\tilde{B}. \quad \text{(C4)}$$

The simulated heat capacity of independent dimers [Fig. 11(a)] exhibits a broad peak at $k_{\rm B}T = 0.35J$ when $B = 0$. The temperature of this maximum is initially suppressed as magnetic field is applied. For $\tilde{B} > 0.7J$, two distinct anomalies become apparent: (i) a low-temperature peak which moves to $T = 0$ as the energy gap is closed [see the energy-level diagram in Fig. 3(c)]; and (ii) a broader feature that moves to progressively higher temperatures as $\tilde{B}$ increases. When $\tilde{B} = J$, the heat capacity exhibits a single maximum at $k_{\rm B}T = 0.61J$.

For $\tilde{B} \geq J$ [Fig. 11(b)], a low-temperature peak in $C_{\rm mag}(T)$ is recovered when the energy gap reopens. This feature tracks to higher temperatures as the field is raised, catching up to and then merging with the high-temperature Schottky anomaly at $\tilde{B} = 1.5J$. The heat capacity $C_{\rm mag}(T)$ then exhibits a single broad peak, whose amplitude and temperature both become greater as $\tilde{B}$ increases further.

For $\tilde{B} \neq J$, the high-$T$ limit of $S_{\rm mag}$ approaches $R\ln 4$ per dimer [Fig. 11(c)]. This is consistent with the change from a populated singlet ground state at low temperatures to the thermal population of all four spin states at higher temperatures. If, however, $\tilde{B} = J$, there is a doubly degenerate ground state and the entropy change per dimer is consequently only $R\ln 4 - R\ln 2 = \frac{1}{2}R\ln 4$.

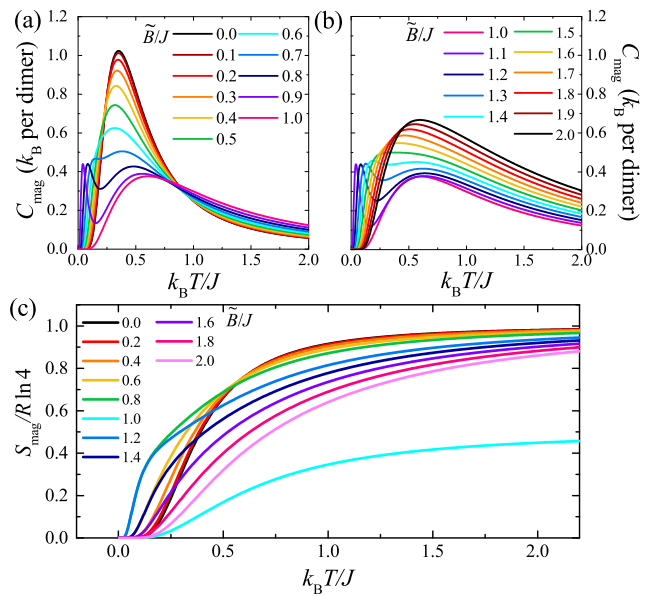

FIG. 11. Simulated heat capacity for independent dimers using Equations (C3) and (C4) for (a) $\tilde{B} \leq J$ and (b) $\tilde{B} \geq J$, where $\tilde{B} = g\mu_{\rm B}B$. (c) Calculated entropy (in units of $R\ln 4$) for independent dimers and selected fields in the range $0 \leq \tilde{B}/J \leq 2$.

The calculated entropy of the system of independent dimers is plotted across the $\tilde{B} - T$ plane in Fig. 12(a).

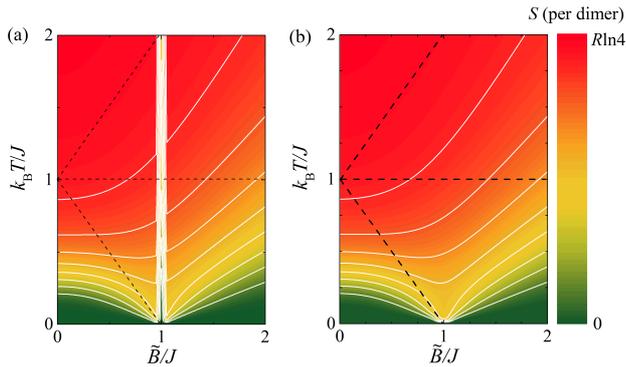

FIG. 12. (a) Calculated $S_{\rm mag}$ per dimer (colormap, in units of $R\ln 4$) across the $\tilde{B}-T$ plane ($\tilde{B}=g\mu_{\rm B}B$) for a system of independent dimers, deduced by integrating the heat capacity data as $C/T$ from Fig. 11. Solid lines are constant entropy contours. Dashed lines are the energies of the dimer spin states. (b) Entropy in the $\tilde{B}-T$ plane, using the heat capacity data in the restricted field ranges of $\tilde{B}\le 0.9J$ and $\tilde{B}\ge 1.1J$ only.

Since the change in $S_{\rm mag}$ on cooling for $\tilde{B}=J$ is half of the full entropy available, the constant entropy contours (isentropes, white lines) exhibit a large spike at this critical field. In real systems, small perturbations will prevent a degenerate ground state from being perfectly realised. Consequently, the large spike in the isentropes is unlikely to be observed. To remove this artefact in the data, the contours in Fig. 12(b) are constructed by using the simulated heat capacity for $\tilde{B}\le 0.9J$ and $\tilde{B}\ge 1.1J$ only, then interpolating $S_{\rm mag}$ over the critical field $\tilde{B}=J$.

### 2. Two-dimensional dispersion

For exchange-coupled dimers, the excited triplet states develop a dispersion relation. This is found by considering the geometry of the interdimer exchange interactions [Fig. 13(a)], and is included in the simulation of the heat capacity using the transformation $J \to J + 2J'\cos k_x \cos k_y$ in Eq. (C2) [here, $\boldsymbol{k}=(k_x,k_y)$ is the dimensionless triplon wavevector]. An average energy is then determined from

$$\langle U \rangle = \frac{\sum_{k_x}\sum_{k_y} U(k_x,k_y)\Delta k_x \Delta k_y}{\sum_{k_x}\sum_{k_y}\Delta k_x \Delta k_y}, \quad (C5)$$

where the discrete sampling of $\boldsymbol{k}$-states runs over $101\times 101$ evenly spaced values spanning the band of excited states [Fig. 13(b)]. Each $k_i$ ($i=x,y$) value is considered in the range $[-\pi,\pi]$. The heat capacity is found numerically from $C_{\rm mag}={\rm d}\langle U\rangle/{\rm d}T$ and the simulation results with $J'=0.28J$ are discussed in the main text [Fig. 6(e)]. Isentropes for this model are constructed following the procedure outlined in the previous section.

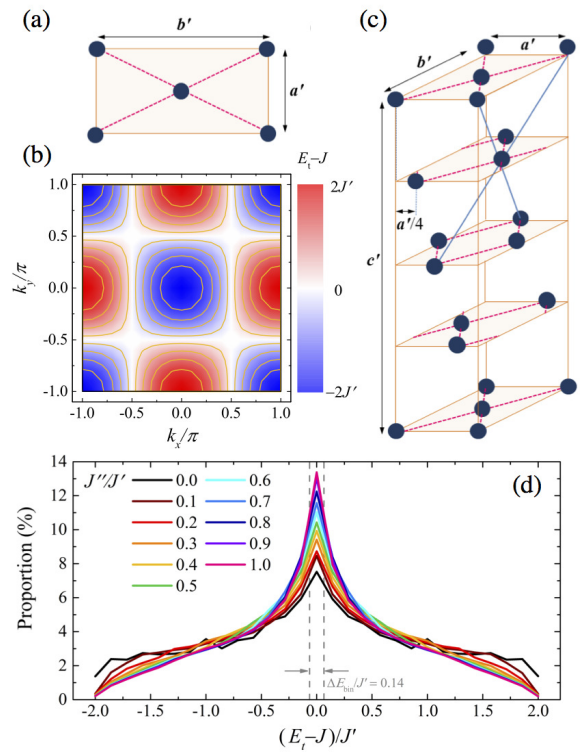

FIG. 13. (a) Schematic geometry of the intraplane exchange network. Dimer centers (black circles) are separated by $\pm(\boldsymbol{a}'/2)\pm(\boldsymbol{b}'/2)$. (b) The energy of the dispersive triplon states $E_{\rm t} = J + 2J'\cos k_x \cos k_y$, in the $k_x-k_y$ plane. Solid lines are constant energy contours. (c) Schematic of the staggered interlayer stacking of dimers. The interlayer interactions between dimer centers act along the vectors $\pm(\boldsymbol{a}'/4)\mp(\boldsymbol{c}'/4)+(\boldsymbol{b}'/2)$ and $\pm(\boldsymbol{a}'/4)\mp(\boldsymbol{c}'/4)-(\boldsymbol{b}'/2)$. (d) Histograms of the proportion of states sampled as a function of the triplon energy ($E_{\rm t}$) in the calculation of the heat capacity using the 3D dispersion. Here, the critical fields (Eq. C6) were fixed and interlayer interactions strengths ($J''/J'$) were considered in the range 0 to 1. The dashed lines show the width of the energy bins ($\Delta E_{\rm bin}=0.14J'$) used in the construction of the histograms. Each curve approximates to the density of states for the triplon excitations for a particular $J''/J'$ ratio.

### 3. Three-dimensional dispersion

Given the geometry of the interlayer interactions [Fig. 13(c)], a three-dimensional dispersion for the triplon excitations can be considered by using the transformation $J \to J + 2J'\cos k_x \cos k_y + 2J''\cos(k_y/2)\cos(k_x/4-k_z/4)$ in Eq. C2. The critical fields of the system now occur at

$$\tilde{B}_{\rm c1,2} = J \mp 2(J'+J''). \quad (C6)$$

We initially consider a dimer network with $J'=0.28J$ and $J''=0$ and study the effects on the heat capacity when the ratio $J''/J'$ is raised from 0 to 1, while the critical fields of the system remain fixed. In each calculation, the energy of the system was found by averaging over unique combinations of discretely sampled $\bm{k}$-states. Here, each $k_i$ ($i = x, y, z$) value was taken at 31 evenly spaced values from $[-\pi, \pi]$.

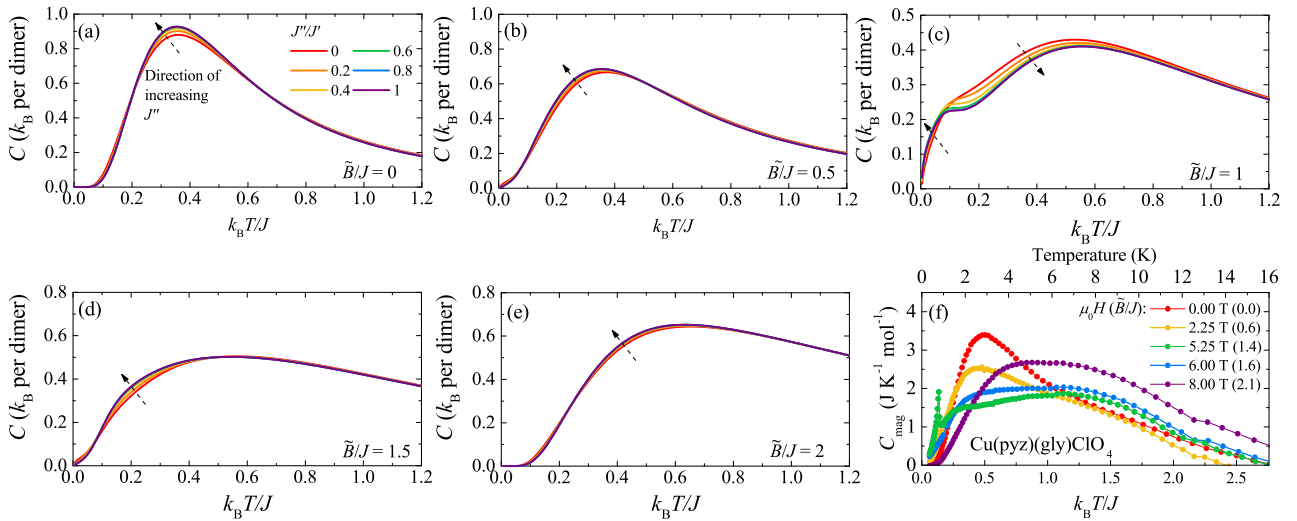

FIG. 14. Simulated heat capacity for a network of interacting dimers with a three-dimensional dispersion relation. Data are shown for fixed applied fields: (a) $\tilde{B}/J = 0$; (b) $\tilde{B}/J = 0.5$; (c) $\tilde{B}/J = 1$; (d) $\tilde{B}/J = 1.5$; and (e) $\tilde{B}/J = 2$, where $\tilde{B} = g\mu_B B$. Colors indicate data calculated at different interlayer interaction strengths in the range $0 \leq J''/J' \leq 1$ [legend in panel (a)] and dashed arrows indicate the trends of features in the data in the direction of increasing $J''$. (f) Measured heat capacity of Cu(pyz)(gly)ClO$_4$. Colors indicate data collected at different applied fields, $\mu_0 H$, with the approximate value of $\tilde{B}/J$ given in brackets.

Histograms of the proportion of excited states sampled at each energy within the triplon bands [Fig 13(d)] show that dispersionless triplons are the most frequently sampled states in the simulations for all values of $J''/J'$ considered. The main effect of raising $J''$ is to increase the proportion of these dispersionless states in the bands.

The simulated $C_{\text{mag}}(T)$ data for the three-dimensional model exhibits a single broad maximum in zero field [Fig. 14(a)]. As $J''/J'$ is raised, the effect of decreasing the proportion of states at the triplon-band extrema [Fig. 13(d)] is to push this broad feature to smaller values of $k_B T/J$. When $\tilde{B}/J = 0.5$, the energy gap to the singlet state is reduced and the height of the broad maximum is suppressed [Fig. 14(b)]. The effects of a greater interlayer interaction at this field are qualitatively similar to the zero-field case.

For $\tilde{B}/J = 1$, $C_{\text{mag}}(T)$ for purely two-dimensional networks ($J'' = 0$) exhibits a Schottky anomaly and an additional shoulder at lower temperatures [Fig. 14(c)]. Introducing $J''$ enhances the temperature of the Schottky anomaly peak. As a result, the low-temperature shoulder becomes better distinguished and starts to be resolvable as an independent peak when the interdimer interactions reach $J'' = J'$. For greater values of $\tilde{B}$ [Fig. 14(d), (e)], the general trend is for changes in $S_{\text{mag}}$ to occur at slightly lower temperatures as the value of $J''/J'$ is enhanced, yielding Schottky anomalies in the heat capacity that are observed at smaller values of $k_B T/J$.

The measured heat capacity of Cu(pyz)(gly)ClO$_4$ [reproduced in Fig. 14(f)] shows field-induced $\lambda$-peaks bounding an $XY$ ordered phase. The occurrence of a long-range ordered phase supports the conclusion that finite interlayer interactions exist in this material. For $\tilde{B}/J = 1.4$ and 1.6, a shoulder in the measured heat capacity at temperatures below the main Schottky anomaly is evident in the data. However, this feature is not so pronounced that it can be identified as a separate peak. By comparing this result to the calculated $C_{\text{mag}}(T)$ traces at similar magnetic fields [Fig. 14(c)], the occurrence of a small shoulder-like feature in the heat capacity of the real material suggests the data are consistent with a relatively small value of $J''/J'$ in the real system.

Lastly, we note the heat-capacity model for dimers with a two-dimensional model underestimates the strength of the MCE observed in the real material (see main text). The effect of including $J''$ in the heat-capacity calculation does not sufficiently increase the temperature of the Schottky anomalies across the studied field range to account for this discrepancy. This implies that, despite the good agreement between data and simulation, there are additional contributions to the spin entropy of Cu(pyz)(gly)ClO$_4$ that are not accounted for by our simple thermodynamic model.

---

* J.D.Brambleby@warwick.ac.uk

† P.Goddard@warwick.ac.uk